\documentclass[twocolumn,floatfix,pra,aps,showpacs]{revtex4}
\usepackage{epsfig,graphicx}
\usepackage{flafter}
\usepackage{amsmath}
\usepackage{color}
\usepackage{braket}
\usepackage[dvipdfm, colorlinks,linkcolor=black,anchorcolor=black,citecolor=black,CJKbookmarks=true]{hyperref}
\begin{document}

\title{First and second sound in a highly elongated   Fermi gas at unitarity}
\author{ Yan-Hua Hou$^{1}$, Lev P. Pitaevskii$^{1,2}$, Sandro Stringari$^{1}$}
\affiliation{1 Dipartimento di Fisica, Universit\`{a} di Trento
and INO-CNR BEC Center, I-38123 Povo, Italy\\
2 Kapitza Institute for Physical Problems RAS, Kosygina
2, 119334 Moscow, Russia}

\date{\today}
\begin{abstract}
We consider a Fermi gas at unitarity trapped  by a highly elongated harmonic potential and solve the equations of two fluid hydrodynamics at finite temperature. The propagation of sound waves as well as the discretized solutions in the presence of weak axial trapping are considered. The relevant thermodynamic functions entering the hydrodynamic equations are discussed in the  superfluid and normal regimes  in terms of universal scaling functions.  Both  first sound and second sound solutions are calculated as a function of temperature and the role of the superfluid density is explicitly pointed out. The density fluctuations in the second sound wave are found to be large enough to be measured as a consequence of the finite thermal expansion coefficient of the gas. Emphasis is given to the comparison with recent experimental data.

\end{abstract}

\maketitle

\section{Introduction}

Propagation of sound is one of the most exciting features exhibited by interacting many-body systems. It provides crucial information on the dynamic behaviour of the system as well as on key thermodynamic quantities. The propagation of sound is particularly interesting in superfluids where two-fluid hydrodynamic theory predicts the occurrence of two different sounds \cite{Tisza40,Landau41}: first sound, which is basically an iso-entropic wave and whose velocity is controlled by the adiabatic compressibility,  and second sound, which corresponds to an isobaric wave where the normal and superfluid components oscillate with opposite phase. Second sound has attracted much attention in the literature of superfluids mainly because the velocity of this peculiar sound is determined by the superfluid density. Actually in liquid $^{4}$He the most accurate determination  of the temperature dependence of the superfluid density is obtained through the measurement of the second sound velocity \cite{Peshkov}. These  data allowed Landau to establish the correct form of the spectrum of the elementary excitations of $^{4}$He, including the roton minimum \cite{Landau47}.

In ultracold atomic gases, below the critical temperature for superfluidity, the propagation of sound has been the object of extensive  theoretical and experimental work in the recent years. A peculiar feature of ultra-cold gases is that they are  confined, the confinement being often of   harmonic shape, which causes the discretization of  sound waves in the form of collective oscillations. At the same time, in the case of highly elongated configurations, it is possible to investigate directly also the propagation of sound waves, by generating a perturbation in the center of the trap and subsequently investigating the time and space propagation of the resulting signal along the long axis \cite{mit,thomas,ueda}.

Most of theoretical investigations in trapped atomic gases have been so far carried out at zero temperature, where only the first sound oscillations exist, both in Bose-Einstein condensed gas and in interacting Fermi gases. The agreement between theory and experiment is pretty good (for a review see, for example, \cite{FGLS} and \cite{GLS}). In the case of BEC gases these studies have permitted to check the validity of  superfluid hydrodynamic theory at zero temperature \cite{stringari96}. In the case of interacting  Fermi gases, along the BEC-BCS crossover, they have permitted to investigate fine details caused by the interactions and quantum statistical effects in the excitation spectrum of the discretized oscillations \cite{stringari04, Astrakharchik05}, including the beyond mean field Lee-Huang-Yang effect \cite{grimmLHY}. The propagation of sound near zero temperature, and in particular the value of the sound velocity, have also been the object of systematic studies, and the general agreement between theory \cite{zaremba,capuzzi} and experiments \cite{mit,thomas} is satisfying, both for Bose and Fermi  superfluids.

Much less is known about the behavior of sound at finite temperature. This is due to various reasons. On  one hand the realization of the collisional regime in the thermal component of a Bose gas, needed to apply the equations of hydrodynamics, is not obvious due to the very dilute and weakly interacting nature of the system. On the other hand the accurate control of  temperature requires sophisticated  experimental techniques. From the theoretical point of view the implementation of dynamic theories at finite temperature is much more difficult than at zero temperature \cite{GriffinBook}, especially in the realistic case  of trapped configurations. The situation becomes particularly challenging in the study of second sound, due to the intrinsic difficulties in generating and  monitoring temperature waves and in providing accurate theoretical predictions for their behavior.

Attempts to excite the relative motion between the condensate and the thermal components of a harmonically trapped Bose gas were first carried out in \cite{oldmit} and more recently in \cite{utrecht} where the use of denser samples permitted to explore collisional damping effects. The propagation of second sound waves in dilute Bose gases was explored in \cite{utrecht2}.
In dilute Bose gas, however, second sound behaves quite differently from the case of strongly interacting superfluids, like $^{4}He$. In dilute Bose gases second sound actually reduces to the oscillation of the condensate in agreement with Tisza's original idea except at very low temperature in the phonon regime \cite{LevSandro}. In a strongly interacting  superfluid it instead corresponds to the oscillation of the gas of elementary excitations relative to the superfluid component, and its observation consequently  gives important information on the excitation spectrum.

Differently from dilute Bose gases, Fermi gases at unitarity (where atoms occupying  different spin states interact with a value of the scattering length much larger than the interparticle distance) behave like strongly interacting fluids and, in this respect, are more similar to liquid $^{4}He$, despite the different statistics. These systems actually exhibit novel and important features and  are characterized by a universal thermodynamic behavior  whose experimental determination has been the object of recent systematic studies \cite{ueda,Nascimbène,MarkMartain} and whose theoretical investigation  represents a stimulating challenge of high interdisciplinary interest \cite{GLS}. Collisions in the unitary Fermi gas are more effective than in dilute Bose gases, so that the non superfluid component of the system can easily achieve the hydrodynamic regime in a wide range of temperatures \cite{grimmHD}.  Furthermore, due to the small compressibility, the superfluid and normal components strongly overlap in space also in the presence of external harmonic confinement, thereby favoring the propagation of second sound.

The achievement of the hydrodynamic regime, as well as an efficient experimental excitation and investigation of sound waves is favored by the use of highly elongated traps.  The discretized  low frequency oscillations of first sound nature in these elongated configurations have been recently measured with high precision in the case of the unitary Fermi gas as a function of temperature \cite{joint, grimmexp}, exploiting fine details of its thermodynamic behavior.  The excitation and observation of wave packets propagating along the long (axial) direction has been also recently carried out at finite temperature \cite{ueda, IBKTN2}.  From the theoretical side the use of highly elongated configurations allows for an important simplification  of the formalism, through the formulation of the so called 1D hydrodynamic approach \cite{Gianluca} whose derivation and implementation in the case of the unitary Fermi gas represent the main goal of the present paper. In this formalism the system preserves the applicability of the local density approximation along the axial as well as the radial directions, but exhibits a typical 1D like behavior in the sense that the fluctuations of the relevant thermodynamic functions (such as the temperature, the axial velocity and the chemical potential) depend uniquely on the axial coordinate \cite{1D}. The 1D like behavior of the low energy mode is the consequence of the effects of viscosity and thermal conductivity  and is favored by a tight radial confinement. It is well suited to describe relevant experimental scenarios as we will discuss in the following sections.

Our paper is organized as follows: In Sec. II we discuss the main features of the 1D hydrodynamic equations and the general conditions required for their applicability. In Sec. III we focus on the case of the unitary Fermi gas for which we derive the relevant 3D and 1D thermodynamic functions  needed  to solve the hydrodynamic equations. In Sec. IV we discuss the basic features of  the variational approach which is implemented in Sec. V and VI, where we provide results for the  first and second sound solutions, respectively. In particular in Sec. V we calculate the discretized frequencies which have been recently measured in \cite{joint,grimmexp}, while in Sec. VI we discuss the behavior of the  second sound solutions both in the cylindrical geometry, where they take the form of sound waves, and in the case of axially trapped configurations, where the eigen-frequencies are discretized. Finally in Sec. VII we draw our conclusions.

\section{Two-fluid hydrodynamics}

In this paper we consider atomic gases at finite temperature confined by a harmonic potential of the form
\begin{equation}
V_{ext}= \frac{1}{2}m\omega^{2}_{\perp}r^{2}_{\perp}+\frac{1}{2}m\omega^{2}_{z}z^{2}
\label{Vho}
\end{equation}
and we will assume highly elongated configurations with trapping frequency satisfying the condition $\omega_z \ll \omega_\perp$, while $m$ is the atomic mass. Our aim is to discuss the dynamic behavior of a  trapped Fermi gas of two spin species interacting with infinite scattering length (unitary limit),
both below and above the superfluid transition. We will discuss the propagation of sound in the cylindrical  geometry ($\omega_z=0$) as well as the discretized solutions with frequency of order  $\omega_z$.

In \cite{Gianluca} it was shown that, under suitable conditions of radial trapping,  it is possible to derive simplified 1D hydrodynamic equations starting from the
usual two-fluid Landau hydrodynamic equations defined in 3D \cite{IMK}. The basic point for such a derivation is the requirement that both the normal velocity field  along the long $z$-th  axis
and the temperature oscillations during the propagation of sound do not depend on the radial coordinates. This requirement is justified in the case of tight radial confinement and is a direct consequence of the effects of viscosity and of thermal conductivity. The condition can be formulated in the simple form
\begin{equation}
\eta \gg mn_{n1}\omega
\label{eta}
\end{equation}
where $\eta$ is the shear viscosity, $n_{n1}$ is the 1D normal density,  obtained by radial integration of the 3D normal density,  and $\omega$ is the frequency of the sound (see discussions at the end of Sec. V). An analogous condition holds for the thermal conductivity. In terms of the radial trapping frequency $\omega_\perp$  the 1D hydrodynamic condition (\ref{eta})
can be rewritten in the form $\omega\ll\omega^2_\perp \tau$, where $\tau$ is a typical collisional
 time, hereafter assumed, for sake of simplicity, to characterize  both the effects of viscosity
 and of thermal conductivity \cite{Gianluca}. The above condition should be satisfied together
with the usual hydrodynamic condition $\omega\tau \ll 1$ \cite{violation} and is satisfied by
the low frequency modes of the trapped gas, of order of $\omega_z$, provided the radial
trapping is sufficiently tight.
It is worth noticing that the above 1D condition implies that also the fluctuations of the chemical potential are independent of the
radial variables.  This follows \cite{Gianluca} from the condition of mechanical equilibrium along the radial
direction $\partial_\perp P + n\partial_\perp V_{ext}=0$ and  from the thermodynamic identity
\begin{equation}
dP=sdT+nd\mu
\label{THidentity}
\end{equation}
 where $s$ is the entropy density, $n$ is the particle density and $P$ is the pressure of the gas. Violation of the radial mechanical equilibrium condition would actually  result in frequencies of the order  $\omega_\perp$
 rather than $\omega_z$.

 By radial integration of the 3D hydrodynamic equations, and following the procedure described
 in \cite{Gianluca}, one  obtains the following 1D hydrodynamic equations:
\begin{eqnarray}\label{Eq.continuity}
m\partial_{t}n_{1}+\partial_{z}j_{z}=0
\end{eqnarray}

\begin{eqnarray}\label{Eq.entropy}
\partial_{t}s_{1}+\partial_z(s_{1}v_{n}^{z})=0
\end{eqnarray}

\begin{eqnarray}\label{Eq.vs}
m\partial_{t}v_{s}^{z}=-\partial_{z}(\mu_1(z)+V_{ext}(z))
\end{eqnarray}

\begin{eqnarray}\label{Eq.current}
\partial_{t}j_{z}=-\partial_{z} P_{1}-n_{1}\partial_{z}V_{ext}(z)
\end{eqnarray}
where the terms $n_{1}$, $s_{1}, P_{1}$ are the radial integrals of their 3D counterparts,
namely the particle density, the entropy density and the local pressure, the integration accounting  for the  inhomogeneity
caused by the radial component of the trapping potential (\ref{Vho}).
In the above equations $j_{z} =m(n_{n1}v_{n}^{z}+n_{s1}v_{s}^{z})$
is the current density, $n_{s1}$ and $n_{n1}$  are the superfluid and
normal number densities respectively with $n_1=n_{s1} + n_{n1}$ while $v_{s}^{z}$
and $v_{n}^{z}$ are the corresponding velocity fields. The continuity equation in
 Eq. (\ref{Eq.continuity}) expresses mass conservation. Equation (\ref{Eq.entropy}) shows that
the entropy of the fluid is carried by the normal fluid. In Eq. (\ref{Eq.vs})
$\mu_1=\mu(T,n({\bf r}_\perp=0,z))$ is the chemical potential calculated on the symmetry axis of the trapped gas  and is determined by the equation of
 state  of uniform matter. Eq. (\ref{Eq.vs}) fixes the law for the superfluid velocity,
while Eq. (\ref{Eq.current}) is the 1D version of the Euler equation for the current.
Since we are interested only in the linear solutions, in the above equations we have omitted
terms quadratic in the velocity. Here and in the following we assume that  the system is large enough to safely carry out the radial integral using the local density approximation.

The ingredients needed to solve the two fluid equations require the knowledge of  the equation of state  $\mu(T,n)$ and of the superfluid density. Theoretically, the calculation of the thermodynamic functions of the unitary Fermi gas is a great challenge due to the absence of a small coupling parameter. There are numerous efforts to develop strong-coupling many-body theories for such a system (see \cite{GLS} and \cite{LiuPhyRep} with references therein and \cite{Nikolay, Haussmann, Bulgac08, Forbes, zhang, Houcke, Drut, OWingate}). In order to probe the hydrodynamic behavior of the two fluid hydrodynamic equations the knowledge of thermodynamics both below and above $T_{c}$ is essential. Actually even below $T_c$ the fluid is normal in the peripheral region where it approaches the classical regime.
Due to the  uncertainties of the theoretical calculations in relevant  temperature ranges,
we have chosen the strategy of using, for the equation of state, the data available from
the recent experimental analysis of the MIT team at unitarity \cite{MarkMartain}.
Universality can then be used to build  the thermodynamic functions for all values of $T$
and $n$ (see Sec. III). Actually the experimental MIT data do not cover the whole range
of temperatures and the information on the equation of state can be implemented and completed
at high temperature by making use of the virial expansion (\cite{HoMueller, LiuPhyRep} with references therein) and, at very  low temperature, by calculating explicitly the phonon contribution which is known to give, in superfluids, the  exact behavior as $T\to 0$ \cite{LevSandro}. As far as the superfluid density is concerned, its present theoretical knowledge is rather poor and we will make use of simple ad-hoc parameterizations in order to provide explicit predictions. The recent experimental investigation of second sound, which is particularly sensitive to the behavior of the superfluid density, has provided the first access to this quantity of fundamental interest \cite{IBKTN2}.

\section{Thermodynamics of the  unitary Fermi gas: from 3D to 1D}

In order to derive the relevant 1D thermodynamic quantities needed to solve the hydrodynamic equations (\ref{Eq.continuity}-\ref{Eq.current}) let us first
discuss the thermodynamic behavior of  uniform matter.

\subsection{3D thermodynamic functions}

At unitarity the s-wave scattering length diverges and, in uniform matter, the remaining
length scales are the thermal wavelength
\begin{equation}
\lambda_{T}=\sqrt{2\pi\hbar^{2}/mk_{B}T}
\label{lambda}
\end{equation}
and the inter-particle distance $n^{-1/3}$. For the same reason the energy scales are fixed by the temperature $T$ and by the Fermi temperature
\begin{equation}
T_F={1\over k_B} {\hbar^{2}\over 2m} (3\pi^{2}n)^{2/3}
\label{TF}
\end{equation}
or, in alternative, by the chemical potential $\mu$. It follows that at unitarity
all the thermodynamic functions can be
expressed \cite{Ho} in terms of a universal function $f_p(x)$ depending on the  dimensionless parameter  $x\equiv \mu/k_BT$. This function can be defined in terms of the pressure of the gas as
\begin{equation}
{P\over k_BT} \lambda_T^{3}\equiv f_{p}(x) \; .
\label{Pnx}
\end{equation}
Using the thermodynamic relation $n= (\partial P/\partial \mu)_T$,  the density of the gas can then be written as
\begin{equation}
n\lambda_T^{3} = f_{p}'(x) \equiv f_n(x) \; .
\label{n}
\end{equation}
From Eq. (\ref{n}) one derives the useful expression

\begin{equation}
\frac{T}{T_{F}} =\frac{4\pi}{[3\pi^{2}f_{n}(x)]^{2/3}}
\label{Tfuni}
\end{equation}
 for the ratio between the temperature and the Fermi temperature (\ref{TF}).

In addition to the functions $f_p(x)$ and $f_n(x)$ it is also useful to define the function

\begin{equation}
f_{q}(x) =\int^{x}_{-\infty}dx^{\prime}f_{p}(x^{\prime})
\label{fq}
\end{equation}
which, as we will show soon, enters some relevant 1D thermodynamic quantities.

In terms of $f_n$ and  $f_p$ we can calculate directly the  thermodynamic functions of the uniform Fermi gas at unitarity. For example, using the thermodynamic relation $S=V(\partial P /\partial T)_\mu$ for the entropy, we find
\begin{equation}
\frac{S}{Nk_{B}}=\frac{s}{nk_{B}}=\frac{5}{2}\frac{f_{p}}{f_{n}}-x
\label{SperN}
\end{equation}
while the specific heats at constant volume and pressure become

\begin{equation}
\frac{C_{V}}{Nk_{B}}=\frac{c_{v}}{nk_{B}}=\frac{15}{4}\frac{f_{p}}{f_{n}}-\frac{9}{4}\frac{f_{n}}{f^{\prime}_{n}}
\label{Cv}
\end{equation}

\begin{eqnarray}\label{Cp}
\frac{C_{P}}{Nk_{B}}=\frac{c_{p}}{nk_{B}}=\left(\frac{15}{4}\frac{f_{p}}{f_{n}}-\frac{9}{4}\frac{f_{n}}{f^{\prime}_{n}}\right)   \frac{5}{3}\frac{f'_{n}f_p}{f^{2}_{n}} \; .
\end{eqnarray}
According to thermodynamics the ratio between $C_P$ and $C_V$ coincides  with the ratio between the  isothermal ($\kappa_T$)  and the adiabatic ($\kappa_S$) depressibilities
\begin{eqnarray}\label{ratio}
\frac{C_{P}}{C_{V}}=\frac{\kappa_T}{\kappa_S}=\frac{5}{3}\frac{f'_{n}f_p}{f^{2}_{n}}
\end{eqnarray}
with  $\kappa_T$ and $\kappa_S$ given,  respectively, by :
\begin{eqnarray}\label{isothermalk}
\kappa_{T}=\frac{1}{n}\left(\frac{\partial n}{\partial P}\right)_{T,N}=\frac{\lambda^{3}f^{\prime}_{n}}{k_{B}Tf^{2}_{n}}
\end{eqnarray}

\begin{eqnarray}\label{adiabatick}
\kappa_{s}=\frac{1}{n}\left(\frac{\partial n}{\partial P}\right)_{S,N}=\frac{3}{5}\frac{\lambda^{3}}{k_{B}Tf_{p}}
\end{eqnarray}
In Eqs. (\ref{SperN}-\ref{Cp}) we have introduced  the entropy ($s$) and the specific heat ($c_v$, $c_P$) densities.

The scaling  function $f_p(x)$ (and  hence
the various thermodynamic functions) can be determined through microscopic many-body calculations
or extracted directly from experiments. In Fig. \ref{figmu} and Fig. \ref{figfnfp}  we show, respectively, the equation of state $\mu/k_BT$ as a function of  $T/T_{F}$
and  the universal  functions $f_{n}(x)$ and $f_{p}(x)$ as a function of $x$,
determined according to the procedures discussed in the following sections. In Fig. \ref{figSC3D}
we show the relevant thermodynamic functions $S/Nk_{B}$, $C_V/Nk_{B}$ and $C_P/Nk_{B}$ as a function of $T/T_F$.

Differently from the previous thermodynamic   quantities,  the superfluid density $n_{s}$ instead requires the knowledge of another independent function. According to dimensional arguments,  at unitarity $n_s$ can be written in terms of a universal function $f_{s}(x)$ as

\begin{eqnarray}\label{sfden}
n_{s}(T,x)=\frac{1}{\lambda^{3}_{T}}f_{s}(x)
\end{eqnarray}
 Its behavior is  known  at low temperature, in the phonon regime (see Sec. III C) \cite{salasnich}, and near the critical point where  it  is predicted  to vanish  as $n_{s} \propto (1-T/T_c) ^{2/3}$ \cite{Josephson}. Here $T_c$ is the critical temperature for superfluidity which, at unitarity can be written in the form
\begin{equation}
T_{c}=\alpha T_{F}
\label{alpha}
\end{equation}
with $\alpha$ a dimensionless universal number.

\subsection{1D thermodynamic functions}

Starting from the above 3D thermodynamic quantities one can calculate the relevant 1D
quantities entering the hydrodynamic equations (\ref{Eq.continuity}-\ref{Eq.current}), whose solution is the main goal of the present paper. In the presence of radial harmonic trapping the
 chemical potential varies along the radial direction according to the law $\mu(r_\perp) =\mu_1 -(1/2)m\omega^2_\perp r^{2}_\perp$, predicted by the the local density approximation, so that one can easily reduce the radial integrals to integrals in the variable $x$.  The following results hold for the 1D density, pressure, entropy and specific heats per particle:
\begin{eqnarray}\label{Eq.1Dden}
n_{1}(x_1,T)=\int dr_{\perp} 2\pi r_{\perp}n=\frac{2\pi}{m\omega_{\perp}^{2}}\frac{k_{B}T}{\lambda_T^{3}}f_{p}(x_1)
\end{eqnarray}

\begin{eqnarray}\label{Eq.1DPress}
P_{1}(x_1,T)=\int dr_{\perp} 2\pi r_{\perp}P=
\frac{2\pi}{m\omega_{\perp}^{2}}\frac{(k_{B}T)^{2}}{\lambda_T^{3}}f_q(x_1)
\end{eqnarray}

\begin{eqnarray}\label{Eq.1Ds}
\frac{s_{1}(x_1,T)}{k_{B}}=\int dr_{\perp} 2\pi r_{\perp}s=\frac{2\pi}{m\omega_{\perp}^{2}}\frac{k_{B}T}{\lambda_T^{3}}\left[\frac{7}{2}f_q(x_1)-x_1f_p(x_1)\right]
\end{eqnarray}

\begin{eqnarray}\label{Eq.1Dcv}
\frac{\bar{c}_{v1}(x_1)}{k_{B}}=T\left(\frac{\partial \bar{s}_{1}/k_{B}}{\partial T}\right)_{n_{1}}=\frac{35}{4}\frac{f_{q}(x_1)}{f_{p}(x_1)}-\frac{25}{4}\frac{f_{p}(x_1)}{f_{n}(x_1)}
\end{eqnarray}

\begin{eqnarray}\label{Eq.1Dcp}
\frac{\bar{c}_{p1}(x_1)}{k_{B}}=T\left(\frac{\partial \bar{s}_{1}/k_{B}}{\partial T}\right)_{P_{1}}
=\bar{c}_{v1}(x_1)\frac{7}{5}\frac{f_{q}(x_1)f_{n}(x_1)}{f^{2}_{p}(x_1)}
\end{eqnarray}
where $x_1= \mu_1/k_BT$ is the value of the chemical potential, in units of $k_{B}T$, calculated on the symmetry axis of the trap, and ${\bar s}_1=s_1/n_1$ is the entropy per particle.

Since ${\bar s}_1$ depends only on the variable $x_{1}$ one finds that the adiabatic derivative of the 1D pressure with respect to the 1D density takes the form
\begin{equation}
\left({\partial P_1 \over \partial n_1}\right)_{\bar{s}_1}={7\over 5}{P_1 \over n_1}
\label{1Dadiab}
\end{equation}
differently from the uniform case where one has, at unitarity,  $(\partial P /\partial n)_{\bar s}=(5/3) P/n$ with $\bar{s}=S/N$. From the comparison between Eq. (\ref{Pnx}) and Eq. (\ref{Eq.1Dden}) one also finds the relationship
\begin{equation}
n_1 = \frac{2\pi}{m\omega_{\perp}^{2}}P({\bf r}_\perp=0)
\label{ho}
\end{equation}
between the 1D density and the pressure calculated at ${\bf r}_\perp=0$. This relationship
holds in the local density approximation for a general fluid  radially confined with  harmonic trapping \cite{hoNature}. It actually follows directly from the radial integration of the general thermodynamic equation  $n= (\partial P/\partial \mu )_T$.

In Fig. \ref{figs1c1} we show the relevant 1D thermodynamic functions calculated as a function of the ratio
\begin{eqnarray}
\frac{T}{T_{F}^{1D}} =\left({\frac{16}{15 \sqrt{\pi} f_p(x_1)} }\right)^{2/5}
\label{TF1}
\end{eqnarray}
where
\begin{equation}
T_F^{1D}=\frac{1}{k_{B}}\left(\frac{15\pi}{8}\right)^{2/5}(\hbar\omega_{\perp})^{4/5}\left(\frac{\hbar^{2}n^{2}_{1}}{2m}\right)^{1/5}
\label{TF1D}
\end{equation}
is the natural definition for the Fermi temperature in 1D cylindrical configurations \cite{Gianluca}. If $n_1$ is calculated for an ideal Fermi gas at zero temperature, $T^{1D}_F$ coincides with the usual 3D definition (\ref{TF}) of Fermi temperature with $n$ calculated on the symmetry axis. If one instead calculates $n_1$ in the unitary Fermi gas at zero temperature one finds the relationship $T^{1D}_F=\xi^{2/5}T_F$ where $\xi$ is the so called Bertsch parameter (see next section).

As concerns the 1D superfluid density, starting from Eq. (\ref{sfden}) we find the expression
\begin{eqnarray}\label{Eq.1Dsfden}
n_{s1}(x_{1},T)=\int dr_{\perp} 2\pi r_{\perp} n_{s}=\frac{2\pi}{m\omega_{\perp}^{2}}\frac{k_{B}T}{\lambda_T^{3}}f_{s1}(x_{1})
\end{eqnarray}

with

\begin{eqnarray}\label{Eq.1Dsfdenn}
f_{s1}(x_{1})=\int^{x_{1}}_{-\infty} dxf_{s}(x).
\end{eqnarray}

From the knowledge of the 1D thermodynamic functions one can also  easily calculate the equilibrium properties in the presence
of axial harmonic trapping, using the local density approximation $\mu_{1}(z)=\mu_{0}-V_{ext}(z)$ for the chemical potential along the $z$-th direction, with $\mu_0$ fixed by the normalization condition $\int dz n_1(z)=N$.
For example the 1D density profile is available from
Eq. (\ref{Eq.1Dden}) by replacing $x_{1}$ with $[\mu_{0}-(1/2)m\omega^2_zz^2]/k_BT$. It is then natural to express the value of $x_0= \mu_0/k_BT$ in terms of the  Fermi temperature $T^{trap}_{F}=(3N)^{1/3}\hbar\bar{\omega}_{ho}/k_{B}$ of the 3D trapped Fermi gas, where $\bar{\omega}_{ho}$ is the geometrical average of the three oscillator frequencies and $N$ is the total number of atoms. One finds
\begin{equation}
T/T^{trap}_{F}=\left(\frac{6}{\sqrt{\pi}}\int^{x_{0}}_{-\infty}dx(x_{0}-x)^{1/2}f_{n}(x)\right)^{-1/3}.
\label{TFtrap}
\end{equation}
This temperature scale will be used to discuss the temperature dependence of the discretized frequencies of the elementary excitations in the presence of $3D$ harmonic confinement.

\subsection{Phonon regime in the  low T limit}

At very low temperatures, corresponding to $T<<T_c$, phonons provide the leading contribution to the thermodynamic behavior of uniform superfluids. In this regime one can easily calculate the relevant thermodynamic functions introduced in the previous section.

Starting from the expression \cite{LevSandro}

\begin{eqnarray}\label{Eq.11}
F=E_{0}-\frac{V\pi^{2}(k_{B}T)^{4}}{90(\hbar c)^{3}}
\end{eqnarray}
for the free energy associated with the phonon excitations in a uniform 3D superfluid, one can easily evaluate the other thermodynamic functions. In the above equation  $E_{0}$ is the ground state energy,  $c$ is the $T=0$ value of the sound velocity,
while $V$ is the volume occupied by the gas. For the unitary Fermi gas one can write
$E_{0}=\frac{3}{5}N\xi k_{B}{T}_{F}$  and  $mc^{2} =\frac{2}{3}\xi k_{B}{T}_{F}$.
Here $\xi$ is the universal Bertsch parameter (\cite{xi},\cite{GLS}),  accounting for the interaction effects of the unitary Fermi gas. Starting from (\ref{Eq.11})  one can calculate the low temperature expansion of the chemical potential $\mu=(\partial F/\partial N)_{T,V}$, pressure $P=-(\partial F /\partial V)_{T,N}$ and entropy $S=-(\partial F /\partial T)_{V,N}$. One finds:

\begin{eqnarray}\label{Eq.12}
\mu=k_{B}T_{F}\left[\xi+\frac{\pi^{4}}{240}\left(\frac{3}{\xi}\right)^{3/2}\left(\frac{T}{T_{F}}\right)^{4}\right]
\end{eqnarray}

\begin{eqnarray}\label{Eq.press}
P=\frac{2}{5}nk_{B}T_{F}\left[\xi+\frac{\pi^{4}}{48}\left(\frac{3}{\xi}\right)^{3/2}\left(\frac{T}{T_{F}}\right)^{4}\right]
\end{eqnarray}

and

\begin{equation}
\frac{S}{Nk_{B}}=\left(\frac{3}{\xi}\right)^{3/2}\frac{\pi^{4}}{60}\left(\frac{T}{T_{F}}\right)^{3}.
\label{entropylowT}
\end{equation}

Using Landau equation for the calculation of the phonon contribution to the normal density one  also finds the result
\begin{equation}
\frac{n_n}{n}=\frac{3\sqrt{3}\pi^{4}}{40\xi^{5/2}}\left(\frac{T}{T_{F}}\right)^{4}.
\label{nnlowT}
\end{equation}

Using definition (\ref{Eq.press}) one can also calculate the large $x$ expansion of $f_p$ and  hence of $f_n$. We find
\begin{eqnarray}\label{Eq.13}
f_{p}(x)=\frac{2}{5}\frac{(4\pi)^{3/2}}{3\pi^{2}}\left[\xi\left(\frac{x}{\xi}\right)^{5/2}
+\frac{\pi^{4}}{96}\left(\frac{3}{x}\right)^{3/2}\right]
\end{eqnarray}
and
\begin{eqnarray}\label{Eq.fnphonon}
f_{n}(x)=\frac{(4\pi)^{3/2}}{3\pi^{2}}\left[\left(\frac{x}{\xi}\right)^{3/2}-\frac{\pi^{4}}{480}\left(\frac{3}{x}\right)^{5/2}\right].
\end{eqnarray}
The first terms in the expansions determine  the $T=0$  value of  the  thermodynamic functions,
 while the second ones account for the first  contribution due to the thermal excitation of phonons.

 An interesting consequence of the expansions (\ref{Eq.13}-\ref{Eq.fnphonon}) concerns the explicit low $T$ behavior of the 1D thermodynamic functions. The 1D entropy (\ref{Eq.1Ds}), as well as the 1D specific heats (\ref{Eq.1Dcv}-\ref{Eq.1Dcp}) and the 1D normal density $n_{n1} = n_1- n_{s1}$  (see (Eq. \ref{Eq.1Dsfden})) exhibit a different $T$ dependence as compared to the corresponding bulk quantities (see Eqs. (\ref{entropylowT}-\ref{nnlowT})). For example the 1D entropy behaves as
\begin{eqnarray}\label{Eq.s1lowT}
\frac{s_{1}(T)}{k_{B}}=\frac{2\pi (k_{B}T)^{5/2}}{m\omega^{2}_{\perp}}\left(\frac{m}{2\pi\hbar^{2}}\right)^{3/2}\gamma
\end{eqnarray}
with  $\gamma=\int^{+\infty}_{-\infty}dx^{\prime}
\left[\frac{5}{2}f_{p}(x^{\prime})-x^{\prime}f_{n}(x^{\prime})\right]$, the integral being convergent since  $(5/2)f_p(x)-xf_n(x)$ decays like $x^{-3/2}$ for large $x$ \cite{gamma}.
Analogously the 1D pressure and the 1D normal density at low temperature behave as
\begin{eqnarray}\label{Eq.P1lowT}
P_{1}(n_1,T)&=&\frac{2}{7}\xi^{3/5}n_1k_{B}T^{1D}_F(n_1)\nonumber\\
&&+\frac{4\pi (k_{B}T)^{7/2}}{7m\omega^{2}_{\perp}}\left(\frac{m}{2\pi\hbar^{2}}\right)^{3/2}\gamma
\end{eqnarray}
and
\begin{eqnarray}\label{Eq.1DnormaldenlowT}
n_{n1}(T)=\frac{2\pi (k_{B}T)^{5/2}}{m\omega^{2}_{\perp}}\left(\frac{m}{2\pi\hbar^{2}}\right)^{3/2}\int^{+\infty}_{-\infty}dx
\nu_{n}(x)
\end{eqnarray}
with the quantity $\nu_{n}(x)$ vanishing as $\frac{\pi^{7/2}}{45}\left(\frac{3}{x}\right)^{5/2}$ at large $x$, in the phonon regime. The above equations reveal that in order to determine the coefficient of the $T^{5/2}$ law in the entropy (\ref{Eq.s1lowT}) and the normal density (\ref{Eq.1DnormaldenlowT}) the knowledge of the functions $f_n$, $f_p$ and $f_q$ are needed for all values of $x$. This is physically due to the fact that in the radial integration the whole temperature range (and not only the large $x$ phonon region) enters the calculation.

It is finally interesting to calculate the low temperature expansion of the 1D chemical potential.
Equation (\ref{ho}) relates the 1D density to the pressure of the gas calculated on the symmetry axis so that the equation of state $\mu_1(n_1,T)$  corresponds to the equation of state of uniform matter  as a function of $P$ and $T$. We can consequently  employ Eqs. (\ref{Eq.12}-\ref{Eq.press}) to derive the low T expansion
\begin{equation}
\mu_1(n_1,T) =k_{B}T^{1D}_{F}\left[\xi^{3/5}-\frac{\sqrt{3}\pi^{4}}{80\xi^{3/10}}\left(\frac{T}{T^{1D}_{F}}\right)^4\right].
\label{mu1T}
\end{equation}
It is worth pointing out that, differently from the case of Eqs. (\ref{Eq.s1lowT}-\ref{Eq.P1lowT}), the first contribution due to thermal effects to $\mu_1$  is  determined by the phonon contribution and exhibits the typical $T^4$  dependence.  It is also interesting to notice the opposite sign exhibited by the thermal correction with respect to the bulk Eq. (\ref{Eq.12}) which implies that  these 1D-like configurations will exhibit a different
thermo-mechanical effect as compared to uniform gases \cite{fountain}.
\subsection{Virial expansion at high T}
The high T behavior of the thermodynamic functions is determined by the virial
expansion which takes the form of an expansion in terms of the fugacity  ($e^{x}\ll1$) and
applies to very large and negative values of $x$. The function $f_p(x)$ can be expanded as:
\begin{eqnarray}\label{Eq.14}
f_{p}(x)=2(b_1e^{x}+b_{2}e^{2x}+...)
\end{eqnarray}
where  $b_j$  are the so-called virial coefficients and the factor '2' comes from  spin degeneracy. The value  $b_1=1$ is fixed by the classical equation of state, while theoretical calculations have provided the values $b_{2}=\frac{3\sqrt{2}}{8}$ \cite{Uhlenbeck} and  $b_{3}=-0.29$ \cite{XHD09} for the second and third coefficients respectively. These values are
consistent with the measurement of the equation of state at high temperature \cite{ueda, Nascimbène, MarkMartain}.

By taking the derivative of the pressure with respect to $x$, we obtain, for the function $f_n(x)$,
the expansion:
\begin{eqnarray}\label{Eq.15}
 f_{n}(x)=2(b_1e^{x}+2b_{2}e^{2x}+...).
\end{eqnarray}

\subsection{The intermediate temperature regime}
\begin{figure}[t]
 \includegraphics[width=0.98\columnwidth]{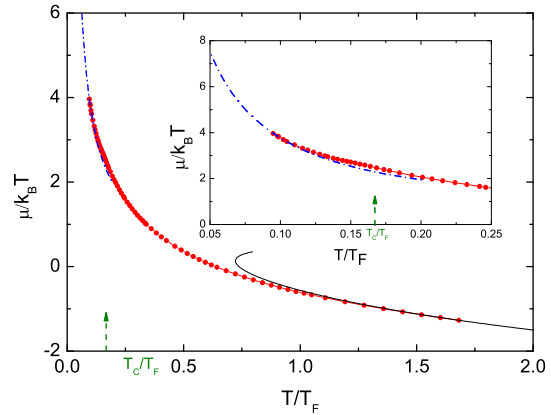}
 \caption{ Equation of state $\mu/k_BT$ versus $T/T_F$. The  blue dash-dotted line corresponds to the phonon contribution to thermodynamics (Sec. III C). The red filled circles correspond to the experiment data in higher $T$ regime, while the black solid line to the virial expansion in classical limit(Sec. III D). The green arrow indicates the critical point
     $T_{c}/T_{F}=0.167(13)$. The inset on the upper right corner is an amplification in the lower $T$ regime. }
  \label{figmu}
\end{figure}

\begin{figure}[t]
 \includegraphics[width=0.98\columnwidth]{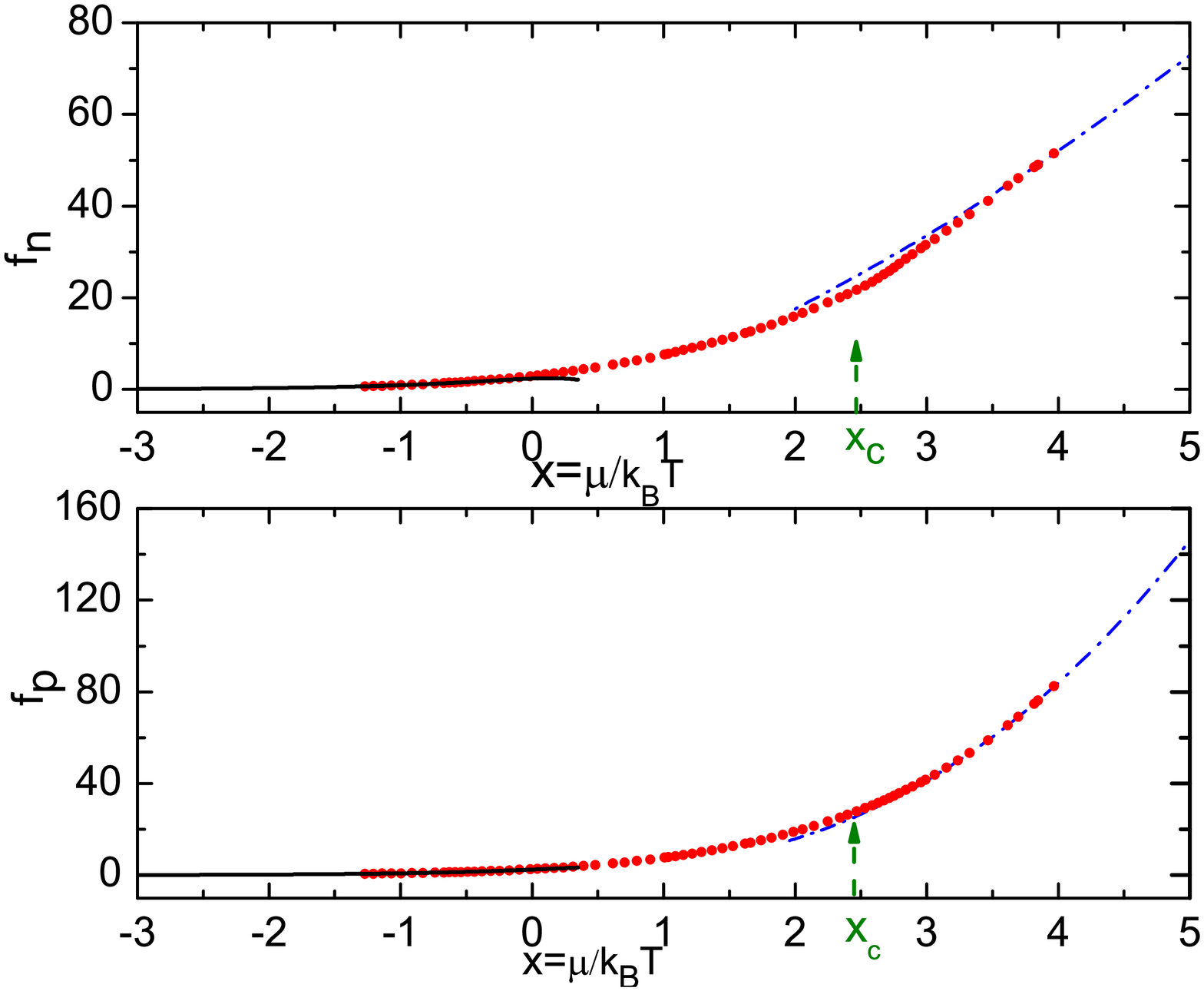}
 \caption{ Universal scaling functions $f_n$ and $f_p$ as a function of the dimensionless variable $\mu/k_BT$. See Fig. \ref{figmu} for the notation.}
  \label{figfnfp}
\end{figure}

\begin{figure}[t]
 \includegraphics[width=0.98\columnwidth]{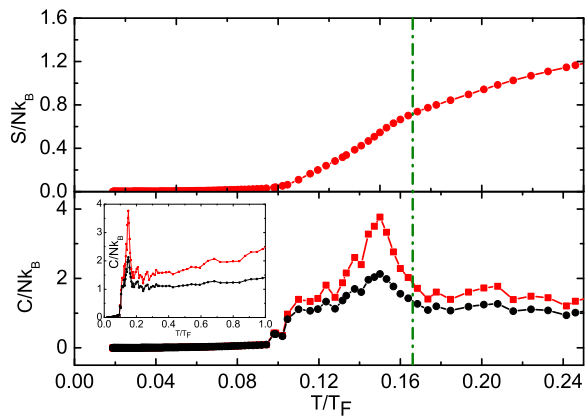}
 \caption{Entropy and specific heats per particle in uniform matter, evaluated using Eqs. (\ref{SperN}-\ref{Cp}). In the lower panel, the red square-guided line corresponds to  $C_{p}/Nk_{B}$; the black full-circle-guided line to $C_{v}/Nk_{B}$; the inset is for the specific heats in a large temperature interval. The vertical green line indicates the critical temperature.}
  \label{figSC3D}
\end{figure}

Through high-precision measurements of the local compressibility, density,
and pressure, the MIT team measured the universal thermodynamic behavior
of the unitary Fermi gas with high accuracy both below and above the
critical temperature for superfluidity overcoming, in particular, the
problem of the direct measurement of the temperature of the gas
\cite{MarkMartain}. These measurements provide an important benchmark for
many-body calculations at finite temperature applied to this strongly
interacting system. They have permitted, in particular, to identify the
superfluid phase transition at the temperature $T_c=\alpha T_{F}$ with
$\alpha= 0.167(13)$, corresponding to the value $x_c= \mu_c/k_BT
 =2.48$.
Also the relevant Bertsch parameter $\xi$, giving the ground state energy in units of the ideal Fermi gas value, was determined with high accuracy ( $\xi= 0.376(4)$). These values are in good
agreement with the best theoretical predictions based on accurate
many-body calculations \cite{Nikolay, Haussmann, zhang,OWingate}. It is worth stressing that in \cite{MarkMartain} the experimental value of the critical temperature was identified by
exploring the peaked structure exhibited by the specific heat at constant
volume near the transition (see Fig. \ref{figSC3D}), by taking explicitly into account
finite resolution effects. This yields a value of $T_c$ slightly higher
than the value where the measured specific heat exhibits its maximum. Figure \ref{figSC3D} shows that the peak exhibited by the specific heat at constant pressure is even
more pronounced, in agreement with the general behavior of the specific
heats near a second order phase transition \cite{LL}. The matching between
the values extracted using the MIT data and the predictions provided by
phonon thermodynamics of Sec. III C is also reasonably good, especially as
concerns the 1D thermodynamic quantities.  At high temperatures these
experiments also confirm with high accuracy the validity of the virial
expansion (see Fig. \ref{figmu} and Fig. \ref{figfnfp}) so that, for the
goals of the present paper, the equation of state of the unitary Fermi gas
can be considered known with reasonably good accuracy at all temperatures.
In the following we will adopt the MIT equation of state, together with
the low and high temperature behavior discussed in Sec. III C and D, to
implement the calculation of the frequency of the collective oscillations
and of the sound velocities within the hydrodynamic formalism.

\begin{figure}[t]
 \includegraphics[width=0.98\columnwidth]{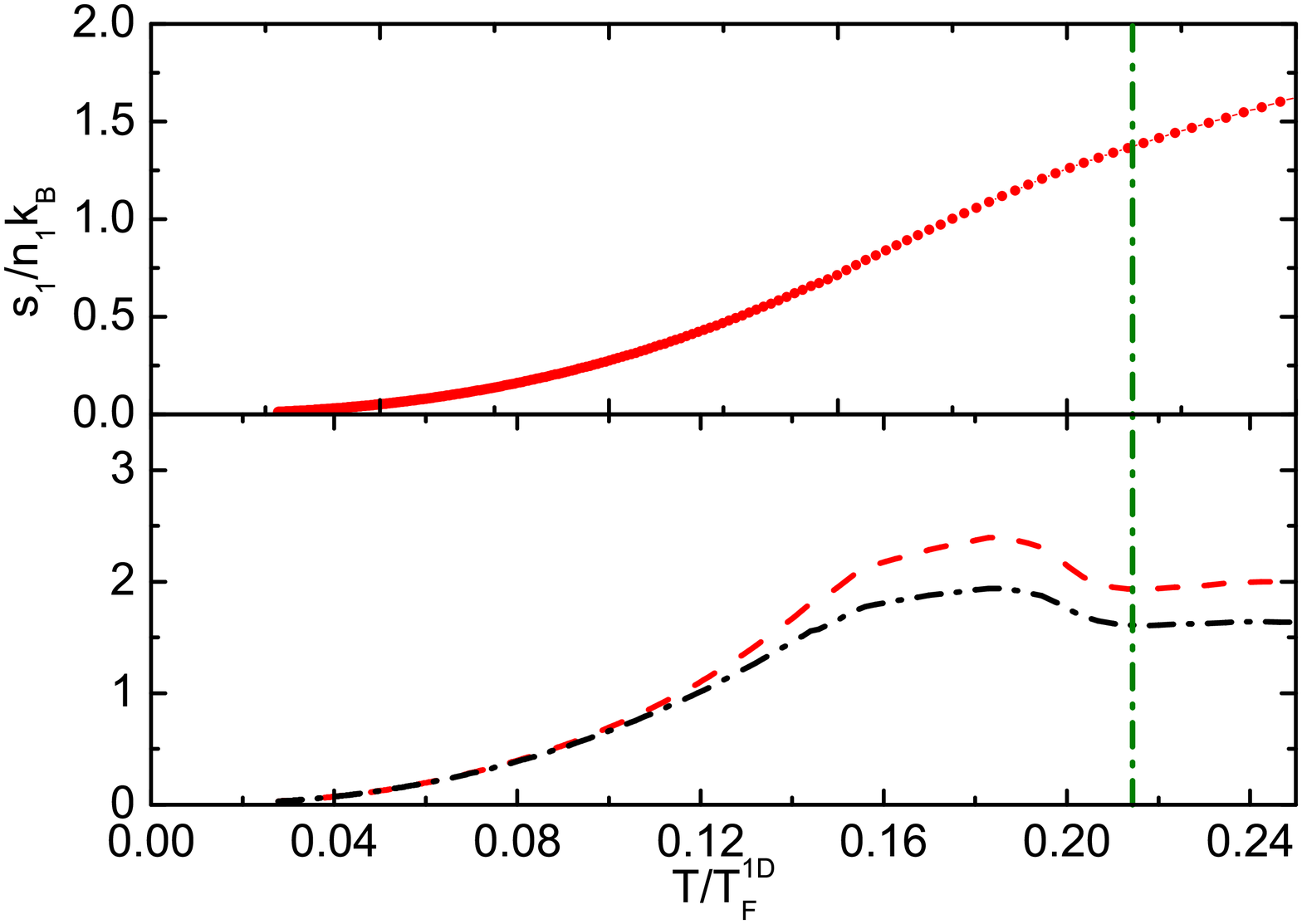}
 \caption{1D entropy and specific heats  evaluated using Eqs. (\ref{Eq.1Ds}-\ref{Eq.1Dcp}). In the lower panel, the red dashed line corresponds to $\bar{c}_{p1}/k_{B}$; the black dash-dotted line to  $\bar{c}_{v1}/k_{B}$. The vertical green line indicates the critical temperature.}
  \label{figs1c1}
\end{figure}

\section{1D variational equations}

The frequencies $\omega$ corresponding to the  solutions  of the  two-fluid hydrodynamic
equations (\ref{Eq.continuity}-\ref{Eq.current}) with time dependence proportional to $e^{-i\omega t}$
can be derived  using  the variational procedure
\begin{equation}
\delta\omega^2/\delta u_n=\delta\omega^2/\delta u_s=0
\label{variational}
\end{equation}
where  \cite{Huathesis}
\begin{eqnarray}\label{Eq.16}
\omega^{2}&=&(\int dz[\frac{1}{n_{1}}\left(\frac{\partial P_{1}}{\partial n_{1}}\right)_{\bar{s}_{1}}
(\delta n_{1})^{2}+2n_{1}\left(\frac{\partial T}{\partial n_{1}}\right)_{\bar{s}_{1}}
\delta n_{1}\delta\bar{s}_{1}+\nonumber\\
& &n_{1}\left(\frac{\partial T}{\partial \bar{s}_{1}}\right)_{n_{1}}
(\delta\bar{s}_{1})^{2}])/\int dzm\left[n_{s1}u_{s}^{2}+n_{n1}u_{n}^{2}\right]
\end{eqnarray}
and $u_{s}$, $u_{n}$ are the displacement field of the superfluid and the normal fluid,
 related to the velocity fields via $\dot{u}_{s}=v^{z}_{s}$ and $\dot{u}_{n}=v^{z}_{n}$. This variational procedure is the 1D version of the 3D approach previously developed in \cite {Taylorvar1, Taylorvar2, TaylorHXG09}. Keeping $\bar{s}_{1}$ constant in the derivatives of  Eq. (\ref{Eq.16}) corresponds  to considering 1D isentropic transformations. The density and entropy fluctuations $\delta n_1$ and $\delta \bar{s}_1$  with respect to equilibrium  are given, in terms of the displacement fields $u_n$ and $u_s$, by  $\delta n_{1}=-\partial_{z}[n_{s1}u_{s}
 + n_{n1}u_{n}]$ and  $\delta \bar{s}_{1}=-u_{n}\partial_{z}\bar{s}_{1}
 +(\bar{s}_{1}/n_{1})\partial_{z}
[n_{s1}(u_{s}-u_{n})]$. Equations (\ref{variational}-\ref{Eq.16}) hold in both uniform and trapped configurations, the effect of the trapping potential entering
through the position dependent thermodynamic functions at equilibrium.

While in uniform configurations one can directly solve the hydrodynamic equations (\ref{Eq.continuity}-\ref{Eq.current}), the use of the variational procedure is particularly convenient in trapped configurations where  analytic solutions of the full hydrodynamic equations are not available.

In order to provide a first quantitative solution of the variational  approach, we will make simple assumptions for the displacement fields of the two fluids.
For first sound, we will assume that the two fluids move in phase with equal displacement fields, i.e. $u_{s}=u_{n}=u$. For second sound we will instead make the assumption that the total current, proportional to $u_nn_{n1}+u_sn_{s1}$,
vanishes  during the oscillation. The coupling between the two modes will be discussed in detail in  Sec. VI. Differently  from the case of dilute BEC gases, where the coupling strongly affects the sound velocities in almost the whole temperature domain \cite{LevSandro},  in the case of Fermi gases  the coupling does not introduce important changes in the value of the collective frequencies, but is important because it releases the assumption that the total current  vanishes in second sound, allowing for significant density fluctuations and hence providing important perspectives for its  experimental detection \cite{smallness}.

\section{First sound}

Employing  the first sound ansatz $u_n=u_s\equiv u$ the expression
(\ref{Eq.16}) for the frequency to be used in the  variational calculation takes the simplified form

\begin{eqnarray}\label{Eq.17}
\omega^{2}=\frac{\int dz n_{1}\left(\frac{\partial P_{1}}{\partial n_{1}}\right)_{\bar{s}_{1}}
\left(\frac{\partial u}{\partial z}\right)^{2}}{\int dzm n_{1}u^{2}} + \omega_z^2
\end{eqnarray}
where we have employed the thermodynamic relation $\partial_zP_1=- n_1\partial_zV_{ext}(z)$ holding at equilibrium
(see Eq. (\ref{Eq.current})). In the  unitary Fermi gas, where the 1D thermodynamic relation (\ref{1Dadiab}) holds,
the variational procedure $\delta \omega^2/\delta u=0$  yields the following equation for the
displacement field:
\begin{eqnarray}\label{Eq.18}
m(\omega^{2}-\omega^{2}_{z})u=\frac{7}{5}m\omega^{2}_{z}z\frac{\partial u}{\partial z}-\frac{7}{5}\frac{P_{1}}{n_{1}}\frac{\partial^{2} u}{\partial z^{2}}.
\end{eqnarray}
The above equations  explicitly reveal that the 1D pressure $P_1$ is the relevant thermodynamic quantity for describing  first sound dynamics. Equation ({\ref{Eq.18}) implies that the center of mass oscillation (dipole mode) characterized by the displacement field $u=const$ is independent of the equation of state and that the axial breathing mode ($u\propto z$)  takes the temperature independent value
$\omega=\sqrt{12/5}\omega_{z}$.

 For axially uniform configurations ($\omega_z=0$) Eq. (\ref{Eq.18}) predicts the propagation of sound waves with dispersion $\omega=c_1q$ and
\begin{eqnarray}
mc^{2}_{1} =\frac{7}{5}\frac{P_1}{n_1}.
\label{c1uniform}
\end{eqnarray}
This  differs from the sound velocity in uniform Fermi gases at unitarity, given by $mc^2=(5/3)P/n$, the difference being caused by the presence of the radial trapping which gives rise to a different condition of
adiabaticity. In Fig. \ref{fig1stsound} we show the value of the first sound velocity $c_1$ as a function of $T/T_F^{1D}$ using the thermodynamic results for $P_1/n_1$ discussed in the previous sections. Using the expansion (\ref{Eq.P1lowT}) for the 1D pressure  one finds that at $T=0$ the first sound velocity approaches the value $c_1=\sqrt{\xi^{3/5}(v^{1D}_{F})^{2}/5}=\sqrt{\xi
v^2_F/5}$ where $v^{1D}_{F}=\sqrt{2k_{B}T_F^{1D}/m}$ and $v_{F}=\sqrt{2k_{B}T_F/m}$ are, respectively, the 1D and 3D Fermi velocities. The quenching of the sound velocity by the factor $\sqrt{3/5}$ with respect to the bulk value $\sqrt{\xi v^2_F/3}$ was first pointed out in \cite{capuzzi}, in analogy with a similar behavior exhibited by Bose-Einstein condensed gases \cite{zaremba}. The figure shows that the 1D iso-entropic prediction (\ref{c1uniform}) well agrees with the experimental data \cite{IBKTN2} for the first sound velocities.
\begin{figure}[t]
 \includegraphics[width=0.98\columnwidth]{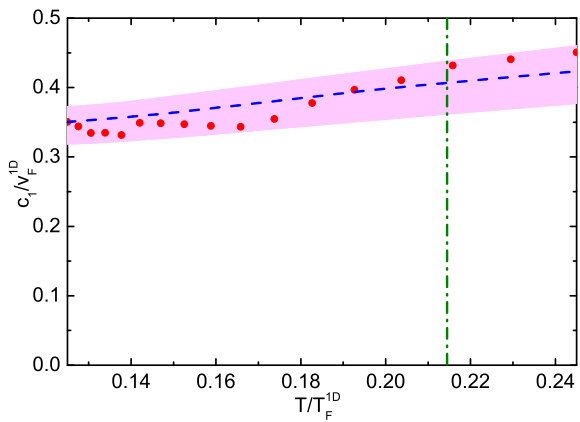}
 \caption{ 1D first sound velocity in units of $v_F^{1D}$ calculated using Eq. (\ref{c1uniform}) (green dashed line). The experiment data (red dots) for the first sound in the shaded region are taken from \cite{IBKTN2}. The shaded area indicates the uncertainty range of experimental data. The vertical green line indicates the critical temperature. }
  \label{fig1stsound}
\end{figure}

Let us now discuss the behavior of the discretized collective oscillations in the presence of axial harmonic trapping. At $T=0$ the 1D pressure  exhibits the position dependence   $P_{1}/n_{1}=(2/7)\left(\mu_{0}-\frac{1}{2}m\omega^{2}_{z}z^{2}\right)$, while
at high temperatures, in the classical limit, one has  $P_{1}/n_{1}=k_{B}T$. In both cases,
it is immediate to find that the solutions of the hydrodynamic equation ({\ref{Eq.18})
are polynomials of the form: $u = a_{k}z^{k} + a_{k-2}z^{k-2} + ...$ with integer values of k.
At zero temperature one  finds the following dispersion relation:

\begin{eqnarray}\label{Eq.19}
\frac{\omega^{2}}{\omega_{z}^{2}}=\frac{1}{5}(k+1)(k+5) \; ,
\end{eqnarray}
while in the high temperature limit one finds
\begin{eqnarray}\label{Eq.20}
\frac{\omega^{2}}{\omega_{z}^{2}}=\frac{7k+5}{5}.
\end{eqnarray}

As expected, Eqs. ({\ref{Eq.19}) and ({\ref{Eq.20}) coincide for $k = 0$
(center of mass oscillation) and $k = 1$ (lowest axial breathing mode), while they predict
different values for the higher nodal solutions. The result for the $k=0$ mode follows from
the universality of the center of mass oscillation in the presence of  harmonic trapping. The fact that the frequency of the  lowest axial breathing oscillation does not depend on temperature is instead a peculiarity of the unitary Fermi gas. It is consistent with the exact scaling solutions exhibited by the two fluid hydrodynamic equations at unitarity \cite{hua2}. The discussion then reveals that only the $k = 2, 3...$ modes are useful in order to explore the effects of the temperature dependence of the equation of state.

In order to provide a simple quantitative prediction for the temperature dependence  of the $k=2$ and $k=3$  frequencies  we develop a variational approach to the solution of the hydrodynamic
equations with the ansatz $u=a_{2}z^2+a_{0}$ and $u=a_{3}z^3+a_{1}z$. This ansatz reproduces exactly the frequencies in both the $T=0$ and high $T$ limits.  Carrying out the variation with respect to the parameters characterizing the displacement fields,
after a straightforward algebra one finds the results

\begin{eqnarray}\label{Eq.22}
\omega^{2}_{k=2}=\frac{129t_{2}-25}{45t_{2}-25}\omega_{z}^{2}
\end{eqnarray}
and
\begin{eqnarray}\label{Eq.28}
\omega^{2}_{k=3}={440 t_3 -252\over 5(25t_3-21)}\omega^{2}_{z}
\end{eqnarray}
for the $k=2$ and $k=3$ frequencies, respectively, where $t_{2}\equiv M_{0}M_{4}/M^{2}_{2}$ and
$t_{3}\equiv M_{2}M_{6}/M^{2}_{4}$ and we have introduced the dimensionless moments
\begin{eqnarray}\label{Eq.moment}
M_{l}=\int_{-\infty}^{x_0} dx(x_{0}-x)^{\frac{l+1}{2}}f_{n}(x)
\end{eqnarray}
where $x_0$ is related to the value of $T/T^{trap}_F$ by Eq. (\ref{TFtrap}). The integrals can be calculated using the data for  the thermodynamic function $f_n(x)$  discussed
in Sec. III which include the proper interpolation between the experimental data from
\cite{MarkMartain}, the low temperature phonon regime as well as  the classical regime, relevant to describe the  low density region on the tails.

The resulting predictions for the temperature dependence of the frequencies are shown in Fig. \ref{figwk2k3},
together with the asymptotic zero temperature and classical values as well as  with the recent experimental results of \cite{joint, grimmexp}.  The results are plotted as a
function of $T/T^{trap}_{F}$ (see Eq. (\ref{TFtrap})). The non monotonic temperature dependence in the higher temperature region is caused by the  presence of the 2nd virial correction into the equation of state. One can also verify \cite{joint} that the variational predictions for the collective frequencies of the $k=2$ and $k=3$ modes are practically indistinguishable from the exact numerical solution of Eq. (\ref{Eq.18}).

Using the same ansatz for the velocity field and the equation of continuity we can also calculate
the density fluctuations of each mode given by $\delta n_{1}=-\partial_{z}[n_{1}u]$
with $\frac{a_{2}}{a_{0}}=-\frac{3}{2}\frac{m\omega^{2}_{z}}{k_{B}T}\frac{M_{0}}{M_{2}}$ and $\frac{a_{3}}{a_{1}}=-\frac{5}{6}\frac{m\omega^{2}_{z}}{k_{B}T}\frac{M_{2}}{M_{4}}$.
The equilibrium density profile $n_1(z)$  is available from Sec. III B, while the moments  $M_\ell$ are given by  Eq. (\ref{Eq.moment}). In Fig. \ref{figmodek2k3} we show the comparison between the theoretical predictions for the $k=2$ and the $k=3$ modes and the corresponding experimental value  at $T=0.1T^{trap}_F$ and $T=0.45T^{trap}_F$ (for the $k=2$ mode) and at $T=0.11T^{trap}_F$ and $T=0.40T^{trap}_F$ (for the $k=3$ mode) \cite{joint,grimmexp}.

The comparison between theory and experiment  is in general
quite satisfying, confirming the validity of the 1D hydrodynamic approach used to predict the temperature dependence  of these  low frequency modes as well as the correctness of the equation of state employed in the calculation of the integrals  $M_\ell$. In particular the relatively small damping shown by experiments in the case of the  $k=2$ mode confirms that the main assumption $\gamma=mn_{1n}\omega/\eta\ll1$ needed to derive the 1D hydrodynamic equations is  reasonably well satisfied. One can estimate the value of the shear viscosity using the data of \cite{Cao}. Then, for the experimental conditions of \cite{joint,IBKTN2,grimmexp}, the parameter $\gamma$ turns out to be of the order of unity. However, a more careful investigation shows the occurrence of a small numerical coefficient in the inequality. One can actually calculate the first correction $\delta c/c$, linear in $\gamma$, to the sound velocity. This correction is imaginary and corresponds to damping. Near $T_c$, employing the notion of "minimal quantum viscosity" \cite{Leclair}, one finds $\delta c/c = -iAmn_1\omega/\eta$, where $\eta$ is the viscosity calculated on the symmetry axis and $A \sim 8.12\times10^{-4}$. In the high $T$ classical limit one instead finds $A \sim 1.62\times10^{-3}$. In both cases the corrections are small in the relevant experimental conditions. The experimental data on the frequency of the $k=3$ mode also reveal a reasonably good agreement with theory, although in this case the observed damping is higher and deviations from theory are observed at higher temperatures \cite{grimmexp}.

\begin{figure}[t]
 \includegraphics[width=0.98\columnwidth]{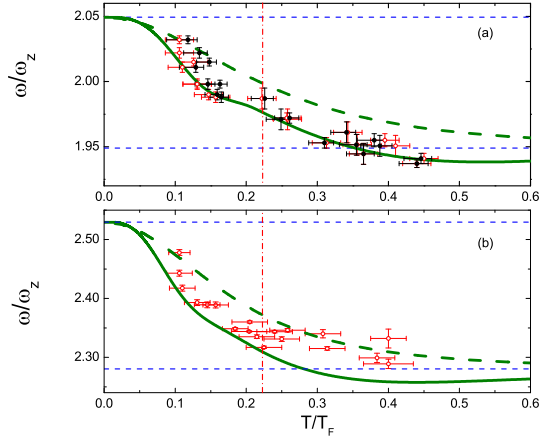}
 \caption{Frequency for the $k=2$ (upper panel) and $k=3$ (lower panel) first sound collective frequencies. Experiment data are from \cite{joint, grimmexp}. The green lines are the theoretical predictions based on Eqs. (\ref{Eq.22}-\ref{Eq.28}) using the equation of state of the unitary (solid) and ideal(dashed) Fermi gas. The thin horizontal dashed lines mark the zero-T superfluid limit (\ref{Eq.19}) and the classical hydrodynamic limit (\ref{Eq.20}), respectively. The red dash-dot vertical lines in (a) and (b) indicate the critical temperature. In this figure and Fig. \ref{figmodek2k3} the Fermi temperature corresponds to the definition $T^{trap}_{F}=(3N)^{1/3}\hbar\bar{\omega}_{ho}/k_{B}$ introduced in the text.}
\label{figwk2k3}
\end{figure}


\begin{figure}[t]
 \includegraphics[width=0.98\columnwidth]{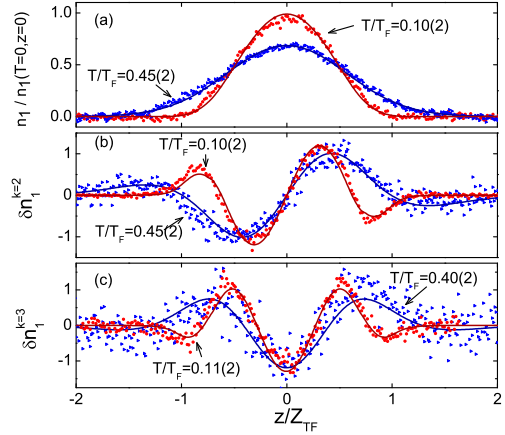}
 \caption{Equilibrium profiles (upper figure) and density oscillations for the $k=2$ (middle figure) and the $k=3$ (lower figure) first sound collective modes at different temperatures from \cite{grimmexp}. See Fig. \ref{figwk2k3} for the definition of the Fermi temperature.}
 \label{figmodek2k3}
\end{figure}

\section{Second sound}
Second sound corresponds to an out-of-phase oscillation of the normal and superfluid components of the fluid. As a first ansatz we assume that the total current be zero
($j_{z}=mn_{n1}v_{n}^{z}+mn_{s1}v_{s}^{z}=0$) which implies that the oscillation corresponds to a pure temperature (or entropy) oscillation,  without any fluctuation of the density.  Differently from first sound the superfluid density $n_s$ plays a crucial role in the propagation of second sound. In superfluid $^4He$ the measurement of the second sound velocity has  actually  provided the accurate determination of $n_s$ as a function of temperature \cite{Peshkov}. At present the theoretical knowledge of $n_s$ is rather poor in the unitary Fermi gas. The first experimental information on $n_s$ has been recently provided by the measurement of the second sound velocity \cite{IBKTN2}. In the following we will make use of simple ansatz for $n_s$ in order to provide a first estimate of the frequency of the second sound oscillations. Under the assumption that the total current vanishes, the expression
(\ref{Eq.16}) for the frequency of second sound  to be used in the  variational calculation takes the simplified form:
\begin{eqnarray}\label{Eq.2ndfre}
\omega^{2}=\frac{\displaystyle\int dz\left(\frac{\partial T}{\partial s_{1}}\right)_{n_{1}} \left[\frac{\partial}{\partial z}\left(\frac{u_{s}s_{1}n_{s1}}{n_{n1}}\right)\right]^{2}}
{\displaystyle\int dz m \frac{n_{s1}n_{1}}{n_{n1}}u^{2}_{s}}
\end{eqnarray}
and the variational condition  $\delta \omega^2/\delta u_{s}=0$  yields the following equation for the
displacement field of the superfluid component:
\begin{eqnarray}\label{Eq.2ndvel}
\omega^{2}u_{s}=-\frac{s_{1}}{mn^{2}_{1}}\frac{\partial}{\partial z}
\left[\left(\frac{\partial T}{\partial \bar{s}_{1}}\right)_{n_{1}}\frac{\partial}{\partial z}\left(\frac{s_{1}n_{s1}u_{s}}{n_{n1}}\right)\right].
\end{eqnarray}
The above equations reveal that the key thermodynamic quantities characterizing the propagation of second sound are the 1D density, entropy, specific heat and superfluid density. The presence of axial trapping is indirectly present through the value of the equilibrium quantities.

From Eq. (\ref{Eq.2ndvel})  one immediately  recovers the second sound velocity for an axially uniform  system by considering
a plane wave solution of the form $e^{iqz}$ for $u_s$. One finds $\omega=c_2q$ with
\begin{eqnarray}\label{Eq.c2}
mc^{2}_{2}=T\frac{\bar{s_{1}}^{2}}{\bar{c}_{v_{1}}}\frac{n_{s1}}{n_{n1}}.
\end{eqnarray}

For uniform configurations it is actually possible to solve exactly the two-fluid hydrodynamic equations (\ref{Eq.continuity}-\ref{Eq.current}) and it is interesting to check the accuracy of the approximate prediction  (\ref{Eq.c2}). Using straightforward thermodynamic relations it is possible to show that  the solutions for the sound velocity emerging from the HD Eqs. (\ref{Eq.continuity}-\ref{Eq.current}) with $V_{ext}(z)=0$ should satisfy the following equation

\begin{eqnarray}\label{soundEq}
&&c^{4}-c^{2}\left[\frac{1}{m}\left(\frac{\partial P_{1}}{\partial n_{1}}\right)_{\bar{s}1}+\frac{1}{m}\frac{n_{s1}T\bar{s}^{2}_{1}}{n_{n1}\bar{c}_{v1}}\right]
\nonumber\\
&&+\frac{1}{m^{2}}\frac{n_{s1}T\bar{s}^{2}_{1}}{n_{n1}\bar{c}_{v1}}\left(\frac{\partial P_{1}}{\partial n_{1}}\right)_{T}=0
\end{eqnarray}
yielding two solutions for the sound velocity, corresponding to the first  ($c_1$) and second  ($c_2$) sound velocities. An accurate  expression for the lower solution (second sound) is derived under the condition
\begin{eqnarray}\label{condition}
\frac{c^{2}_{2}}{c^{2}_{1}}\frac{\bar{c}_{p1}-\bar{c}_{v1}}{\bar{c}_{v1}}\ll1.
\end{eqnarray}
In this case one finds
\begin{eqnarray}\label{Eq.cp2}
mc^{2}_{2}=T\frac{\bar{s_{1}}^{2}}{\bar{c}_{p_{1}}}\frac{n_{s1}}{n_{n1}}
\end{eqnarray}
which differs from Eq. (\ref{Eq.c2}) because of the presence of the specific heat at constant pressure rather than at constant density. The two specific heats actually exhibit a  different behavior for temperatures close to $T_c$ (see Fig. \ref{figs1c1}).  When $T\to 0$ the specific heat at constant pressure and at constant volume coincide and, as a consequence of the temperature dependence of the 1D thermodynamic functions in the low temperature regime (see Sec. III B), the second sound velocity vanishes like $\sqrt{T}$, differently from what happens in the bulk where it  approaches the value $ c_1/\sqrt{3}$ \cite{LevSandro}.  At finite temperature  expression (\ref{Eq.cp2}) is very accurate in reproducing the lower solution of (\ref{soundEq}) for all temperatures, the condition (\ref{condition}) being always well satisfied.  The above discussion then  reveals that second sound can be  regarded as an oscillating wave at constant pressure, rather than at constant density, as previously assumed in the derivation of (\ref{Eq.c2}). This is the consequence of the finite value of the 1D thermal expansion coefficient, $\alpha_{1}=-\frac{1}{n_{1}}\left(\frac{\partial n_{1}}{\partial T}\right)_{p1}$. Differently from the second sound velocity, the velocity of first sound is instead  negligibly affected by the coupling, provided condition (\ref{condition}) is satisfied and is consequently very accurately described by Eq. (\ref{c1uniform}) at all temperatures.

The finite value of the thermal expansion coefficient has the important consequence that the density fluctuations during the propagation of second sound are not negligible \cite{Nikuni,Huietal}. Actually, under the condition $c_2 \ll c_1$, the ratio between the relative density and temperature fluctuations for second sound takes the simple expression
\begin{eqnarray}\label{Eq.denosci}
\frac{\delta n_{1}/n_{1}}{\delta T/T}=\frac{T}{n_{1}}\left(\frac{\partial n_{1}}{\partial T}\right)_{p1} =\frac{5}{2}-\frac{7}{2}\frac{f_{n}f_{q}}{f^{2}_{p}},
\end{eqnarray}
following from the thermodynamic relation $\alpha_{1}=\frac{5}{2T}\frac{\bar{c}_{p1}-\bar{c}_{v1}}{\bar{c}_{v1}}$. The ratio (\ref{Eq.denosci}) turns out to be negative \cite{figGianluca} and is shown in Fig. \ref{figdenosci2nd}. It should be compared with the result $\frac{\delta n_{1}/n_{1}}{\delta T/T}=\frac{T}{n_{1}}\left(\frac{\partial n_{1}}{\partial T}\right)_{\bar{s}1}=\frac{5}{2}$  characterizing the propagation of first sound, where the derivative is  calculated at constant  entropy rather than at constant pressure.  It is remarkable that the ratio (\ref{Eq.denosci}) is significantly large in a useful range of temperatures, thereby  revealing that second sound can be observed by looking at the density fluctuations of the propagating signal \cite{IBKTN2}.

\begin{figure}[t]
 \includegraphics[width=0.98\columnwidth]{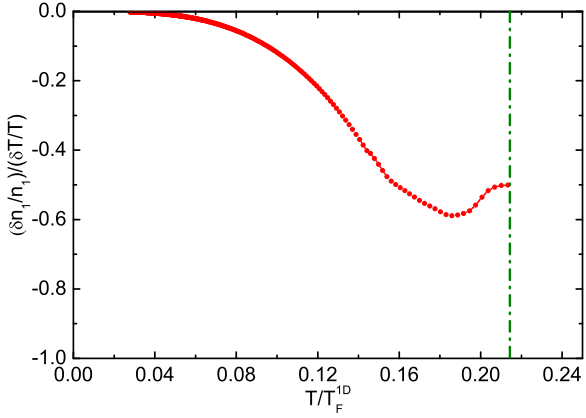}
 \caption{ Ratio (\ref{Eq.denosci}) between the relative density and temperature fluctuations calculated for 1D second sound. The vertical green line indicates the critical temperature.}
  \label{figdenosci2nd}
\end{figure}
In Fig. \ref{fig2ndsound} we show  the prediction for the second sound velocity (\ref{Eq.cp2}) using two different models for $n_s$ (see Fig. \ref{figsfden}). The first model employs the formula $n_s/n=(1-T/T_c)^{2/3}$, accounting for the correct critical exponent $2/3$ characterizing the vanishing of $n_{s}$ near the critical point. A second model employs the phenomenological expression $n_s/n=1-(T/T_c)^{4}$  which also vanishes at $T=T_{c}$ and exhibits, at low temperature the correct $T^4$ behavior, although the coefficient of the $T^4$ law is about $8$ times larger than the one predicted by the phonon  contribution to the normal density (see Eq. (\ref{nnlowT})). The second sound velocity depends in a crucial way on the choice of the model for $n_s$ so that the measurement of $c_2$ is expected to provide useful information on its temperature dependence. The ansatz $n_s/n=1-(T/T_c)^{4}$ provides a better description of the measured data in the relevant temperature regime explored in the recent experiment of \cite{IBKTN2} as explicitly shown by figs. (\ref{fig2ndsound}) and (\ref{figsfden}).

\begin{figure}[t]
 \includegraphics[width=0.98\columnwidth]{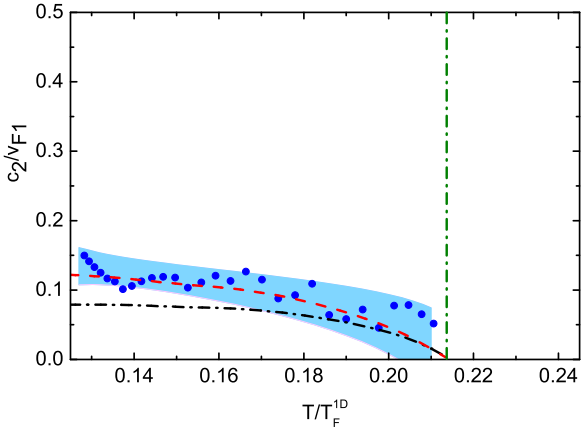}
 \caption{1D second sound velocity  in the axially uniform configuration, calculated using Eq. (\ref{Eq.cp2}) and two different models for the superfluid density: the red dashed-line corresponds to the phenomenological ansatz: $n_{s}/n=1-(T/T_{c})^{4}$ while the black dash-dotted-line to the choice: $n_{s}/n=(1-T/T_{c})^{2/3}$. The scattered blue circles are the experimental data \cite{IBKTN2}. At low temperature the 1D second sound velocity is expected to vanish like $\sqrt{T}$. The shaded area indicates the uncertainty range of experimental data. The vertical green line indicates the critical temperature.}
\label{fig2ndsound}
\end{figure}

\begin{figure}[t]
 \includegraphics[width=0.98\columnwidth]{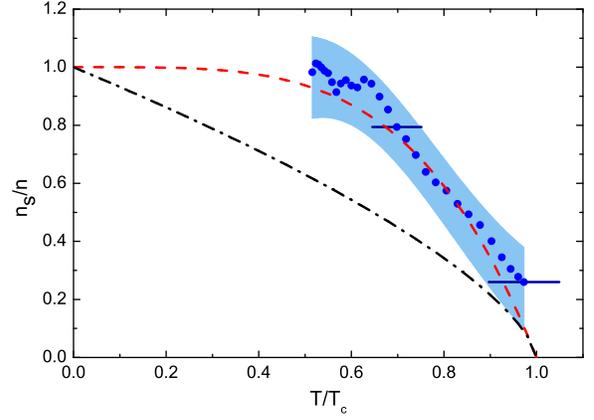}
 \caption{Uniform superfluid density: the red dashed-line corresponds to the phenomenological ansatz: $n_{s}/n=1-(T/T_{c})^{4}$ while the black dash-dotted-line to the choice: $n_{s}/n=(1-T/T_{c})^{2/3}$. The blue dots are the experimental data \cite{IBKTN2}. The shaded area indicates the uncertainty range of experimental data.}
\label{figsfden}
\end{figure}
In the presence of harmonic trapping along the $z$-th direction, the second sound modes are discretized and we use the variational approach (\ref{Eq.2ndfre}) to obtain first estimates for the collective frequencies.  The lowest frequency mode of second sound nature is expected to be of dipolar nature for which we make the simplifying assumption that the displacement field $u_{s}$ of the superfluid component is constant in space and $u_{n}$ is fixed by the condition $n_{s}u_s+n_{n}u_n=0$ of vanishing total current. In Fig. \ref{figomega2} we show the resulting prediction for the temperature dependence of the lowest second sound mode, using the two  models for the superfluid density discussed above.  We have checked that the inclusion of higher order terms in  the  polynomial ansatz for $u_s$ introduces minor corrections (less than $10$ \%).

An important feature of the second sound frequency is that it vanishes when the temperature approaches the critical value. This result  differs from the one  predicted in 3D isotropic
configurations \cite{TaylorHXG09,levin} and can be understood noticing that an estimate for the discretized frequency can be obtained using the expression $\omega \sim c_2q$  with $q\sim1/R_{s,z}$ where $R_{s,z}$ is the size of the superfluid along the $z$-th direction. On the other hand the main temperature dependence of the second sound velocity, as $T\to T_c$,  is given by the the 1D superfluid velocity that behaves
like $\sqrt{n_{s1}} \sim \sqrt{n_sR_{s,\perp}^2}$  and is hence  proportional to the bulk superfluid density calculated in the center of the trap and the size of the superfluid along the radial direction. Since the ratio $R_{s,\perp}/R_{s,z}$ in the LDA is given by $\frac{\omega_{z}}{\omega_{\perp}}$ and $n_s$ vanishes as one approaches the transition temperature, the second sound frequencies will vanish too.

\begin{figure}[t]
 \includegraphics[width=0.98\columnwidth]{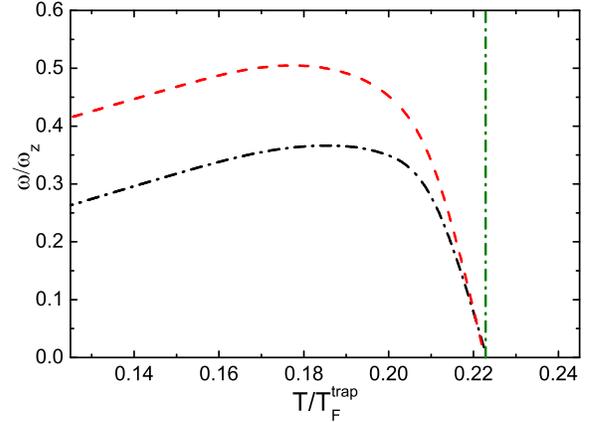}
 \caption{Frequency for the lowest discretized  second sound  mode in an axially trapped configuration with $\omega_z \ll \omega_\perp$. See Fig. \ref{fig2ndsound} for the notations. The vertical green line indicates the critical temperature.}
 \label{figomega2}
\end{figure}

The coupling between the two unperturbed second sound and first sound discussed above and in Sec. V can be estimated using a variational approach. To this purpose we will look for solutions of the variational hydrodynamic equations in the form $u_{s}=a u^{(1)}+ u^{(2)}_{s}$
and $u_{n}=a u^{(1)}+ u^{(2)}_{n}$ for  the superfluid ($u_{s}$) and normal ($u_{n}$) displacement fields, respectively. Here $u^{(1)}\equiv u_{n}^{(1)}=u_s^{(1)}$ corresponds to the velocity field of the first sound solutions discussed in Sec. V, while $u_n^{(2)}$ and $u_s^{(2)}$ are the velocity fields of the uncoupled second sound solutions satisfying the condition of vanishing total current. By inserting the ansatz in (\ref{Eq.16}) we find, after a straightforward calculation, the expression

\begin{eqnarray}\label{Eq.31}
\omega^{2}=\frac{a^{2}\omega^{2}_{1}+\omega^{2}_{2}\frac{K_{2}}{K_{1}}
-a\frac{U_{1,2}}{K_{1}}}{a^{2}+\frac{K_{2}}{K_{1}}}
\end{eqnarray}
for the collective frequency as a function of the variational parameter $a$, where
$K_{1}=(1/2)\int dz mn_{1}u_{1}^{2}$, $K_{2}=(1/2)\int dz m(u_s^{(2)})^2 n_{1}n_{s1}/n_{n1}$ and
$U_{1 ,2}= (2/5)T\int dz
\frac{\partial u_{1}}{\partial z}\frac{\partial s_{1}n_{s1}u^{(2)}_{s}/n_{n1}}{\partial z}
$ and we have used the identity $n_{1}(\frac{\partial T}{\partial n_{1} })_{\bar{s}_{1}}
=\frac{2}{5}T$ holding at unitarity.
By imposing the variational condition $\delta \omega^2 /\delta a =0$ we find the result
\begin{eqnarray}\label{Eq.33}
\omega^{2}=\frac{\omega^{2}_{1}+\omega^{2}_{2}\pm\sqrt{(\omega^{2}_{1}-\omega^{2}_{2})^{2}
+\frac{U^{2}_{1,2}}{K_{1}K_{2}}}}{2}
\end{eqnarray}
for the frequency of the two coupled modes.
When applied to uniform matter, using result (\ref{c1uniform}) and (\ref{Eq.c2}) for the uncoupled first ($\omega_1$) and second ($\omega_2$) sound frequencies, the above procedure reproduces exactly the two decoupled solutions given by roots of Eq. (\ref{soundEq}).
As an  example of application in the presence of harmonic trapping we have considered the coupling between the dipole second sound solution discussed above and the $k=2$ first sound solution
discussed in Sec. V. The $k=2$ mode is actually the lowest frequency first sound mode that can be coupled to the dipole second sound mode, being characterized by the same parity symmetry. The numerical calculation shows that the changes in the value of the second sound frequency caused by the coupling are very small.

Let us finally point out that the discretized second sound oscillations discussed above are expected to be more damped than the first sound ones discussed in the previous section.
The reason is that the thermal conductivity in the normal phase tends to
infinity near the transition point \cite{hohenberg} and is consequently
large near the boundary between the superfluid and the normal phases. This
is expected to result in the penetration of the temperature fluctuations
into the normal phase, resulting in an increase of damping.


\section{Conclusions}

We have provided a systematic discussion of the two fluid hydrodynamic behavior exhibited by the unitary Fermi gas in the presence of a highly elongated harmonic potential. The main achievements contained in the paper are summarize below.

i) We have presented  an exhaustive discussion of the relevant 3D and 1D  thermodynamic functions, like the pressure, the entropy and the specific heats at constant density and at constant pressure, whose knowledge is required in order to solve the hydrodynamic equations. The  thermodynamic functions are identified using the most recent experimental data obtained at MIT, through the introduction of universal scaling functions which emphasize the universality of the unitary Fermi gas. The  matching of  the MIT data with the low T behavior of the 3D thermodynamic functions fixed by the thermal excitation of phonons and  with the high T virial expansion has been explicitly discussed.
Particularly interesting results concern  the behavior of the 1D quantities which are calculated by radial integration of the 3D thermodynamic functions using the Local Density Approximation. The  behavior of the 1D thermodynamic functions at low temperature  is not uniquely fixed by the thermal excitation of phonons as happens in uniform superfluids, but involves also the thermal regimes at higher temperature in the peripheral radial region. Their temperature dependence at low $T$ has been explicitly calculated.

ii) We have solved the 1D hydrodynamic equations derived in \cite{Gianluca}  using a variational formulation of the hydrodynamic equations. Explicit results are given for both the   first and second sound modes. While the first sound solutions are basically determined by the 1D adiabatic compressibility the second sound solutions are sensitive, in addition to the entropy and the specific heat, to the superfluid density of the system, a rather elusive quantity which cannot be determined by the knowledge of the equation of state of the system.

iii) We have provided results for both 1D uniform and axially trapped configurations. In the first case the solutions of the hydrodynamic equations take the form of sound wave whose velocity has been  systematically investigated  for both first and second sound, employing different models for the superfluid density. In the presence of axial trapping the lowest excitations take the form of collective oscillations whose discretized frequencies are calculated as a function of temperature.  The theoretical predictions for the first discretized sound solutions are compared with recent experiments carried out  both below and above the critical temperature for superfluidity \cite{joint}. A systematic discussion of the propagation of first and second sound in 1D uniform configurations was also carried out and a detailed analysis of recent experimental results was presented \cite{IBKTN2}.

iv) An important feature emerging from our studies is that in highly elongated configurations the finite value of the thermal expansion coefficient makes the second sound mode an oscillation at constant 1D pressure, rather than at constant 1D density and an explicit formula for the resulting  density fluctuations has been derived  as a function of temperature. This has the important consequence that, except at very low temperature,  the density fluctuations characterizing second sound are sizable, thereby making this mode observable in experiments.

Open questions to address in future works concern the  damping of  the collective modes  caused by viscosity and thermal conductivity  and a more quantitative check of the applicability of the 1D hydrodynamic description employed in the present paper. The conditions of applicability of this approach, i.e. the independence of the fluctuations of temperature and of the velocity field on the radial coordinate, are actually ensured by  the crucial role played by viscosity, thermal conductivity in the presence of  tight radial trapping.

The authors would like to acknowledge systematic discussions and fruitful collaborations with R. Grimm, M. J. H. Ku, E. R. S$\acute{´a}$nchez Guajardo, L. A. Sidorenkov, M. K. Tey and M. W. Zwierlein. We are grateful to M. J. H. Ku and  M. W. Zwierlein for providing the relevant experimental data characterizing the universal functions of the unitary Fermi gas, systematically employed in the present paper. L. P. Pitaevskii wishes to thank P. Hoheneberg for an insightful discussion. This work has been supported by ERC through the QGBE grant and by Provincia Autonoma di Trento.


\begin{thebibliography}{99}
\bibitem{Tisza40} L. Tisza, J. Phys. Radium {\bf 1}, 164, 350 (1940).

\bibitem{Landau41} L. D. Landau, J. Phys. USSR {\bf 5}, 71 (1941).

\bibitem{Peshkov} V. P. Peshkov, J. Phys. USSR {\bf 8}, 381 (1944); V. P. Peshkov, J. Phys. USSR {\bf 10}, 389 (1946).

\bibitem{Landau47} L. D. Landau, J. Phys. USSR {\bf 11}, 91 (1947).

\bibitem{mit} M. R. Andrews, D. M. Kurn, H.-J. Miesner, D. S. Durfee, C. G. Townsend, S. Inouye, and W. Ketterle, Phys. Rev. Lett. {\bf79}, 553 (1997).

\bibitem{thomas} J. Joseph, B. Clancy, L. Luo, J. Kinast, A. Turlapov, and J. E. Thomas, Phys. Rev. Lett. {\bf98}, 170401 (2007).

\bibitem{ueda}  M. Horikoshi, S. Nakajima, M. Ueda, and T. Mukaiyama, Science {\bf327}, 442 (2010).

\bibitem{FGLS} F. Dalfovo S. Giorgini, L. P. Pitaevskii, and S. Stringari, Rev. Mod. Phys. {\bf71}, 463 (1999).

\bibitem{GLS} S. Giorgini and L. P. Pitaevskii and S. Stringari, Rev. Mod. Phys. {\bf80}, 1215 (2008).

\bibitem{stringari96} S. Stringari, Phys. Rev. Lett. {\bf77}, 2360 (1996).

\bibitem{stringari04} S. Stringari, Europhys. Lett.  {\bf65}, 749 (2004).

\bibitem{Astrakharchik05} G. E. Astrakharchik, R. Combescot, X. Leyronas, and S. Stringari, Phys. Rev. Lett. {\bf95}, 030404 (2005).


\bibitem{grimmLHY} A. Altmeyer S. Riedl, C. Kohstall, M. J. Wright, R. Geursen, M. Bartenstein, C. Chin, J. Hecker Denschlag, and R. Grimm, Phys. Rev. Lett. {\bf98}, 040401 (2007).

\bibitem{zaremba}  E. Zaremba, Phys. Rev. A {\bf57}, 518 (1998).

\bibitem{capuzzi} P. Capuzzi, P. Vignolo, F. Federici, and M. P. Tosi, Phys. Rev. A. {\bf73}, 021603(R) (2006).

\bibitem{GriffinBook} A. Griffin, T. Nikuni, and E. Zaremba, Bose-Condensed Gases at Finite Temperature (Cambridge, 2009).

\bibitem{oldmit} D. M. Stamper-Kurn, H.-J. Miesner, S. Inouye, M. R. Andrews, and W. Ketterle, Phys. Rev. Lett.{\bf81}, 500 (1998).

\bibitem{utrecht} R. Meppelink, S. B. Koller, J. M. Vogels, H. T. C. Stoof, and P. van der Straten, Phys. Rev. Lett. {\bf103}, 265301 (2009).

\bibitem{utrecht2} R. Meppelink, S. B. Koller, and P. van der Straten, Phys. Rev. A {\bf80}, 043605 (2009).

\bibitem{LevSandro} L. P. Pitaevskii and S. Stringari, Bose-Einstein Condensation
  (Oxford, New York, 2003).

\bibitem{Nascimbène} S. Nascimb$\grave{e}$ne, N. Navon, K. J. Jiang, F. Chevy, and C. Salomon, Nature {\bf463}, 1057 (2010).

\bibitem{MarkMartain} M. J. H. Ku, A. T. Sommer, L. W. Cheuk, and M. W. Zwierlein, Science {\bf335}, 563 (2012).

\bibitem{grimmHD} M. J. Wright, S. Riedl, A. Altmeyer, C. Kohstall, E. R. S$\acute{a}$nchez Guajardo, J. Hecker Denschlag, and R. Grimm, Phys. Rev. Lett. {\bf99}, 150403 (2007).

\bibitem{joint}  M. K. Tey, L. A. Sidorenkov, E. R. S$\acute{a}$nchez Guajardo, R. Grimm, M. J. H. Ku, M. W. Zwierlein, Y.-H. Hou, L. Pitaevskii, and S. Stringari, Phys. Rev. Lett. {\bf110}, 055303 (2013).

\bibitem{grimmexp} E. R. S$\acute{a}$nchez Guajardo, M. K. Tey, L. A. Sidorenkov,
and R. Grimm, Phys. Rev. A {\bf87}, 063601 (2013).

\bibitem{IBKTN2}  L. A. Sidorenkov, M. K. Tey, R. Grimm, Y.-H. Hou,	L. Pitaevskii, and S. Stringari, Nature {\bf498}, 78 (2013).

\bibitem{Gianluca} G. Bertaina, L. P. Pitaevskii, and S. Stringari, Phys. Rev. Lett. {\bf105}, 150402 (2010).

\bibitem{1D} The regime we consider here should not be confused with the strict 1D regime where all the particles occupy the lowest single particle state of the radial harmonic potential.

\bibitem{IMK} I. M. Khalatnikov, An Introduction to the Theory of Superfluidity (Benjamin, New York, 1965).

\bibitem{violation} In the opposite $\omega\gg \omega^{2}_{\perp} \tau$ regime  the solutions of the 3D hydrodynamic equations exhibit a quite different behavior \cite{Gianluca}.

\bibitem{LiuPhyRep} X.-J. Liu, arXiv: 1210.2176, Physics Report, to be published.


\bibitem{Nikolay} E. Burovski, N. Prokof$^{,}$ev, B. Svistunov, and M. Troyer, Phys. Rev. Lett. {\bf96}, 160402 (2006).

\bibitem{Haussmann} R. Haussmann, W. Rantner, S. Cerrito, and W. Zwerger, Phys. Rev. A {\bf75}, 023610 (2007).

\bibitem{Bulgac08} A. Bulgac, J. E. Drut, and P. Magierski, Phys. Rev. A {\bf 78}, 023625 (2008).

\bibitem{Forbes} M. M. Forbes, S. Gandolfi, and A. Gezerlis, Phys. Rev. Lett. {\bf 106}, 235303 (2011).





\bibitem{OWingate} O. Goulko and M. Wingate, Phys. Rev. A {\bf82}, 053621 (2010).

\bibitem{zhang} J. Carlson, S. Gandolfi, K. E. Schmidt and S. Zhang,  Phys. Rev. A {\bf 84},  061602(R) (2011).

\bibitem{Houcke} K. Van Houcke,	F. Werner, E. Kozik, N. Prokof$^{,}$ev, B. Svistunov, M. J. H. Ku, A. T. Sommer, L. W. Cheuk, A. Schirotzek, and M. W. Zwierlein, Nat. phys. {\bf8}, 366 (2012).

\bibitem{Drut} J. E. Drut, T. A. L$\ddot{a}$hde, G. Wlazłowski, and P. Magierski, Phys. Rev. A {\bf 85}, 051601(R) (2012).






\bibitem{HoMueller} T.-L. Ho and E. J. Mueller, Phys. Rev. Lett. {\bf92}, 160404 (2004).

\bibitem{Ho} T.-L. Ho,  Phys. Rev. Lett. {\bf92}, 090402 (2004).

\bibitem{salasnich} In a recent paper (Luca Salasnich, Phys. Rev. A {\bf82}, 063619 (2010)) the temperature dependence of the superfluid density was calculated beyond the phonon regime
applying Landau's  equation for $n_n$ in a BEC-BCS famework to account for the effects of the single-particle excitations.

\bibitem{Josephson} B. D. Josephson, Phys. Lett. {\bf21}, 608(1966).

\bibitem{hoNature}  T.-L. Ho and Q. Zhou, Nat. Phys. {\bf6}, 131(2010).

\bibitem{xi} G. F. Bertsch, 1999, in the announcement of the Tenth International Conference on Recent Progress in Many-Body Theories (unpublished).

\bibitem{gamma} Using the values of $f_p$ and $f_n$ determined according to the procedure discussed in Sec. III E we find the value $\gamma=63.2$.

\bibitem{fountain} D. J. Papoular, G. Ferrari, L. P. Pitaevskii, and S. Stringari, Phys. Rev. Lett. {\bf109}, 084501 (2012).

\bibitem{Uhlenbeck} E. Beth and G. E. Uhlenbeck, Physica {\bf4}, 915 (1937).

\bibitem{XHD09} Xia-Ji Liu, Hui Hu  and Peter D Drummond, Phys.Rev.Lett  {\bf 102}, 160401 (2009).

\bibitem{LL} L. D. Landau and E. M. Lifshitz, Statistical Physics, Part
1 (Pergamon Press, Oxford, 1980).

\bibitem{Huathesis} Y.-H. Hou, PhD thesis, in preparation.

\bibitem{Taylorvar1} E. Taylor and A. Griffin, Phys. Rev. A {\bf72}, 053630 (2005).

\bibitem{Taylorvar2} E. Taylor, H. Hu, X.-J. Liu, and A. Griffin, Phys. Rev. A {\bf77}, 033608 (2008).

\bibitem{TaylorHXG09} E. Taylor, H. Hu, X.-J. Liu, L. P. Pitaevskii, A. Griffin, and S. Stringari, Phys. Rev. A {\bf80},  053601 (2009).

\bibitem{smallness} The smallness of the  corrections in the velocity of second sound  due to the coupling is actually compatible with the occurrence of sizable density fluctuations. This can be  understood looking at the equation of continuity which, for a wave propagating with wave vector $q$ and frequency $\omega = c_2q$,  yields the relation $mc_{2}\delta n_{1}=j_z$, explicitly revealing that, since the second sound velocity is small, the smallness of $j_z$ does not necessarily imply the smallness of the fluctuations $\delta n_{1}$ of the 1D density.

\bibitem{hua2} Y.-H. Hou, L. P. Pitaevskii, and S. Stringari, Phys. Rev. A {\bf87}, 033620  (2013).

\bibitem{Cao} C. Cao, E. Elliott, H. Wu and J. E. Thomas, New J. Phys. {\bf13}, 075007 (2011).

\bibitem{Leclair} A. LeClair, New J. Phys. {\bf13}, 055015 (2011).

\bibitem{Nikuni} E. Arahata and T. Nikuni,  Phys. Rev. A {\bf80}, 043613 (2009).

\bibitem{Huietal} H. Hu, E. Taylor, X.-J. Liu, S Stringari, and A Griffin, New J. Phys. {\bf12}, 043040 (2010).

\bibitem{figGianluca} The quantitative differences exhibited by Fig. \ref{figdenosci2nd} with respect to Fig. 2 of \cite{Gianluca} are due to the more accurate thermodynamic ingredients used in the present calculation.

\bibitem{levin} Y. He, Q. Chen, C. C. Chien, and K. Levin, Phys Rev. A {\bf 76}, 051602(R) (2007).

\bibitem{hohenberg} B. Halpern and P. Hohenberg,  Rev. Mod. Phys. {\bf49}, 435 (1977).

\end{thebibliography}
\end{document}